\documentclass[twocolumn,secnumarabic,amssymb, nobibnotes, aps, prd]{revtex4-2}

\usepackage{amsmath}

\newcommand{\be}{\begin{equation}}
\newcommand{\ee}{\end{equation}}
\newcommand{\ba}{\begin{eqnarray}}
\newcommand{\ea}{\end{eqnarray}}

\newcommand{\bal}{\begin{align}}
\newcommand{\eal}{\end{align}}

\newcommand{\nn}{\nonumber}

\begin{document}

\title{Covariant higher order perturbations of branes in curved spacetime }%

\author{Riccardo Capovilla, Giovany Cruz, and Edgar Yair L\'{o}pez}%
\address{
Departamento de F\'{\i}sica, Cinvestav-IPN, Av. Instituto Polit\'ecnico Nacional 2508,
col. San Pedro Zacatenco, 07360, Gustavo A. Madero, Ciudad de M\'exico, M\'exico}
\email{capo@fis.cinvestav.mx, gcruz@fis.cinvestav.mx, elopez@fis.cinvestav.mx}
\date{October 2021}%
\begin{abstract}
The treatment of higher order perturbations of  branes 
is considered using a covariant variational approach. This covariant variational approach brings to the forefront the geometric
structure of the underlying perturbation theory, as opposed to a more commonly used  `direct approach', that ignores the variational origin. In addition, it offers a clear calculational advantage with respect to so called `gauge fixed' treatments that distinguish tangential and normal modes, as it emphasizes the symmetries of the geometric models that
describe the brane dynamics.
 We restrict our attention to a brane action that depends at most on first derivatives of the embedding functions of the worldvolume spanned by the brane in its evolution. We consider first and second variations of the action that describes the brane dynamics. The first variation produces the equations of motion, as is well known. In the second variation we derive the Jacobi equations for these kind of models, and we emphasize the role of the Hessian matrix.
 This is extended to third order in variations, first in a flat and then in a curved spacetime background. 
 Further, we specialize  to the relevant case of the Dirac-Nambu-Goto action that describes extremal branes.
The proper setting of a covariant  variational approach allows to go in principle to geometric models that depend on higher derivatives
of the embedding functions, and higher order perturbations, with the due complications involved, but with a solid framework in place.
 \end{abstract}

\maketitle
%\tableofcontents

\section{Introduction} 
Brane mechanics refers to the study of the dynamics of branes embedded in a higher dimensional space, usually called ambient space, background space, or target space. This is a generalization of the  notion of a relativistic particle,  a 0-brane. Then  we can have strings or 1-branes. The term branes is leaving the dimension arbitrary. Branes  are highly relevant in the description of different physical scenarios where the relevant physical degrees of freedom are confined to an appropriate sub-region of interest.  Branes are described by a local action that is a functional of the geometry of the worldvolume spanned by the brane
in its evolution.  The guiding principles in the construction of the action are reparametrization invariance and
background diffeomorphism invariance. This turns the action into a geometric model.
The term `brane mechanics' was introduced by Carter  \cite{Carter, Carterbrane}, where it is emphasized the range
of applications of the subject. Since then, the range has been widely extended in important ways.

At the classical level, even the simplest geometric models produce equations of motion that are highly non-linear. An important special case are strings propagating in a flat background, where the equations of motion turn out to be linear, in an appropriate
gauge \cite{Zwiebach}. Besides this special case, in general one confronts a serious challenge in the analysis of the
possible solutions of the equations of motion.
In addition, the non-linearity can be made even worse if one considers a non trivial background spacetime,  with non vanishing curvature, see {\it e.g.} \cite{LF94} for the case of a string, where the background curvature plays the role of an external force. In face of these difficulties,  one has to resort
to  perturbation theory.  The first step is to encounter a relevant exact solution
to perturb about. To attain this, symmetries are invoqued and exploited, to make the problem tractable.   The next obvious  step is  a  direct linearization of the equations of motion that produce a set of equations for  field  deviations of fields that (on-shell) reduce to coupled quadratic oscillators, and so on.
This `direct approach' is the one more commonly used in the literature \cite{vilenkin1994cosmic}.
There is also a  `covariant direct approach' that exploits a perturbation taken as a directional covariant derivative along a deviation vector, or deformation, and that is especially advantageous in a curved background, see {\it e.g.} \cite{defos}.

The main subject of this paper is to offer a complete covariant variational approach that does not use any gauge-fixing at any stage, where
by gauge-fixing it is meant a split of brane perturbations in their normal and tangential modes with respect to the worldvolume. This approach
sets brane mechanics as a covariant classical field theory, trying  to avoid any idiosyncratic notation, and using standard familiar language, that hopefully will be useful to the reader. The covariance is with respect to both background diffeomorphisms
and worldvolume reparametrization invariance.  The main tool that we exploit  is a covariant variational derivative, inspired by the
pioneering work of Ba$\dot{\mbox{z}}$a\'{n}ski \cite{Bazpol}, and its higher order extensions.
From this point of view, conservation laws become easily accessible via Noether's theorem, as well as how they behave under perturbations \cite{CapoNoether}. In general, one has at its disposal the whole  weaponry of the calculus of variations, properly generalized to
a field theory. The emphasis on covariance is not only desirable, but needed because gauge invariance is a subtle point in perturbation theory.
As one goes to higher orders in perturbation theory, it becomes more and more difficult to disentangle physical modes from gauge
unphysical modes. As emphasized especially by Carter, explicit covariance allows one to sidestep many artificial complications.
In any case, once covariant expressions are obtained, with reasonable minimal pain, one has always the option to gauge-fix to one heart's desire. In the classical case this is especially important to get to a seed solution of physical interest.
Perturbations then would need to respect the gauge fixing conditions. When considering possible quantum corrections, it is mandatory in order to arrive to a proper definition of physical states.

The action for the brane is defined as an integral over the worldvolume of a Lagrangian density of weight one, to ensure reparametrization
invariance.
Its first order variation produces  the Euler-Lagrange derivative  and a Noether current. The latter is instrumental in the derivation of the
canonical linear momentum, that will play a central role in our approach. The vanishing of the first variation,  under suitable boundary conditions, gives the equations of motion
for the system. The second order variation of the action has been considered by many authors in a `direct approach'. This corresponds to a linearization of the equations of motion \cite{guven1993covariant, garriga1991perturbations, LF94}. An approach related to ours has been offered  in the important work by Battye and Carter \cite{BC00}.  The motivation
to go to second order is manifold, see {\it e.g.} \cite{Mars} for hypersurfaces in arbitrary spacetimes. In brane mechanics, the second order variation is used to study
stability of solutions, see {\it e.g.} \cite{Lovelock} for a covariant study of perturbations in Lovelock type brane gravity. In addition, like the geodesic deviation equation for a relativistic particle, the vanishing of the second variation, on-shell, produces
Jacobi equations  that describe the relative motion of neighboring branes connected by a deviation vector. In a curved background,
they take the form of a kinetic term that involves the Hessian for the model, with the background Riemann curvature
playing the role of an external force.
The second variation also gives the index of the geometric model. The index is of paramount importance in various
geometrical contexts, including geometry in the large, see {\it e.g.} \cite{FT} for minimal surfaces. In addition, in the study of quantum or thermal fluctuations
the second variation appears as the `shape operator' that in a path integral determines  both one-loop contributions and
the semi-classical regime.

In many instances, second order variations is not enough, as
degeneracies may emerge, and a third variation of the action is needed. In any case, it is useful to establish a pattern in
perturbation theory that potentially can be extended to higher orders. This should be of interest especially in relativistic
astrophysical applications, where observational data are becoming more and more precise, and a covariant perturbation
theory is a needed theoretical tool. In this context, the works \cite{1+3fT,1+3fR} are a recent attempt to match upcoming cosmological data with perturbative theoretical models beyond General Relativity.
The third variation we derive exhibits  an interesting structure, with a kinetic term that
depends on the Jacobi operator, inherited from the second variation, acting on the second order perturbations, and a source term
that depends on the known first order perturbations. This structure was known previously in classical field theory, see   \cite{Bazpol}.
One previous study on second order perturbation of Dirac-Nambu-Goto [DNG] branes in a curved background
by Kiosses and Nicolaidis needs to be mentioned in this context \cite{KN14} (see also \cite{LA01}). As far as we know, it was the first attempt
to go to a description of second order perturbations in full generality. Their strategy is a `direct approach' splitting
the perturbations in parts tangential and normal to the worldvolume. Although extremely convenient at first order, where
tangential deformations can be associated to a mere reparametrization, and ignored safely, at higher order things
become quite complicated, and  the risk is to miss the forest for the trees. A covariant variational approach allows to understand
the geometric structure of perturbation theory. In addition, we consider any codimension for the brane, although codimension one,
for a hypersurface and codimension two are simpler to deal with. The presence of source terms that depend not only on the background
Riemann curvature, but also on its covariant derivative is identified.

In recent years, extremal surfaces and their perturbations have emerged as actors in the context of AdS/CFT or Gauge/Gravity duality. In \cite{RT06} it was conjectured that the entanglement entropy of a spatial subregion in a CFT can be calculated perturbatively using the area of a codimension-2 spacelike minimal surface in the dual gravity theory. A covariant version of this holographic entanglement entropy  for arbitrary regions and `extremal' surfaces was given in \cite{CovRT07}. Based on this, the authors of \cite{Bao1,Bao2} argued that the Jacobi operator for 2-dimensional surfaces in a particular coordinate system is uniquely fixed using only boundary data. Classically in the gravity side, holographic 
entanglement entropy is proportional to a surface area. However, probably this is not the whole story.  One may include quantum fields adding to the entanglement, that in turn transform the extremal surfaces into `quantum extremal surfaces'. They are defined as stationary points of Bekenstein's generalized entropy \cite{netta}.
Branes have also been considered for the physics of the large. One imagines the observable universe as being a brane embedded in a higher dimensional place, with all of its fields and particles constrained to the brane, only gravity is free to `leak' into the bulk at high energies. These brane-world models offer possible phenomenological corrections to General Relativity \cite{MK, regge2016general}.\\

There are other contexts, unrelated, where a brane geometrical description is useful and has been proved very successful both theoretically and in its experimental confirmation in the description of equilibrium configurations of lipid membranes in soft matter physics, see {\it
e.g.}  \cite{Deserno} and references therein. There is a common geometrical language, but with respect to more speculative relativistic settings, there
is also the crucial advantage of experimental verification, to test theoretical endeavour.
A geometric approach to fluid lipid membranes in space, using a covariant variational approach is proposed in \cite{CapoSimLipid,CapoBending}. Hydrodynamics on curved surfaces can be studied with a geometric description of submanifolds \cite{FluidMembranes20}. In this context, the theoretical work by Armas and collaborators is worth mentioning,{\it e.g.}  in \cite{Armas}, a thorough study of brane and edge dynamics is given.  \\

This paper is organized as follows. Section \ref{Notation} gives a brief  overview of the embedding  geometry and sets our notation.
Section \ref{GeneralVar} establishes a general covariant variational approach for relativistic brane models that depend at most on first order derivatives on the embedding functions. We construct systematically the first, second and third order variations of the action, by considering covariant variations with respect to a one parameter family of embeddings. For the sake of clarity,
the third order variation is performed first  in a flat background, then we add the complication of a curved background. Section \label{DNGbrane} applies the general formalism to the case of the DNG action.
Finally, in Section \ref{Discussion} we offer a brief discussion.\\

\section{Embedding Geometry} \label{Notation}

In this section, we describe briefly the geometry of the embedding of a time-like submanifold, or worldvolume,
that depicts the evolution of a relativistic object, or brane, in a fixed background spacetime.

The semi-Riemannian background spacetime is  $\{ M , g_{\mu\nu} \} $ of dimension $N+1$ with local coordinates $x^\mu$ 
($\mu, \nu, \dots = 0, 1, \dots, N$), and endowed with a metric $g_{\mu\nu}$ of signature with mostly plus signs. Spacetime indices are lowered and raised with the 
spacetime metric $g_{\mu\nu}$ and its inverse $g^{\mu\nu}$. The background spacetime covariant derivative  $\nabla_\mu$ is assumed to be torsionless and compatible with the background metric $g_{\mu\nu}$,  $\nabla_\rho g_{\mu\nu} = 0$, with Riemann curvature tensor defined as
$[ \nabla_\mu , \nabla_\nu ] V^\rho =  - R_{\mu\nu\sigma}{}^\rho V^\sigma$, where we follow the conventions of \cite{Wald}.

The brane of dimension $d$ sweeps  in its evolution a worldvolume $w$ of dimension $d+1$. The worldvolume $w$ can be
described in a parametric form  by the  (time-like) local embedding functions $x^\mu =  X^\mu (\xi^a)$,
 in $\{ M , g_{\mu\nu} \} $, where $\xi^a$ are local coordinates for $w$ 
 ($  a, b, \dots = 0, 1, \dots, d$). The $d+1$ tangent vectors to the worldvolume  $w$ are
  $X^\mu_a  = \partial_a X^\mu = \partial X^\mu / \partial \xi^a $, and the induced metric, or {\textit{first fundamental form}}, on $w$ is
\be
\gamma_{ab} = g_{\mu\nu} X^\mu_a X^\nu_b\,,
\label{IndMetricDef}
\ee
with determinant $\gamma$.  The time-like character of the wordvolume is assured by requiring $\gamma < 0$.  
 Tangential indices are raised with the inverse induced metric $\gamma^{ab}$ and lowered with $\gamma_{ab}$.
 We denote with $\overline{\nabla}_a$ the worldvolume torsion-less  covariant derivative compatible with the induced metric $\gamma_{ab}$,
 $\overline{\nabla}_a \gamma_{bc} = 0$, with Riemann curvature $ {\mathcal R}_{abc}{}^d $. 

The extrinsic geometry of the worldvolume can be introduced via the spacelike normal vectors $n^{\mu}{}_i$  to the
worldvolume  $(i,j \dots = 1, 2, \dots, N-d)$ that are defined up to a sign and a rotation
 by the orthogonality condition $ g_{\mu\nu}  X^\mu_a n^\nu{}_i = 0$ and normalized with
$g_{\mu\nu} n^\mu{}_i n^\nu{}_j = \delta_{ij}$, where $\delta_{ij}$ is the Kronecker delta. We leave the co-dimension arbitrary. Of course, there are noticeable simplifications for a hypersurface, $N-d = 1$. We also note the completeness relationship
\be
g^{\mu\nu} = h^{\mu\nu} + n^{\mu\nu} =  \gamma^{ab} X^\mu_a X^\nu_b + n^\mu{}_i n^{\nu\, i} \,,
\label{eq:comple}
\ee
that defines the tangential and normal projectors $h^{\mu\nu}$ and $n^{\mu\nu}$, respectively.

The generalization of the classical Gauss-Weingarten equations for submanifolds to a curved background is
given by
\ba
\nabla_a X_b^\mu &=& \gamma_{ab}{}^c X_c^\mu - K_{ab}{}^i n^\mu_i \,, \label{eq:gw1}\\
\widetilde\nabla_a n^\mu{}_i &=& K_{ac\, i} \gamma^{cb} X_b^\mu \label{eq:gw2} \,. 
\ea
Here $\gamma_{ab}{}^c $ is the Christoffel symbol associated with the worldvolume covariant derivative $\overline{\nabla}_a$,
and $\nabla_a=X^\nu_a\nabla_\nu$ is the projection of the background spacetime covariant derivative onto the worldvolume.
The extrinsic curvature tensor, or {\textit{second fundamental form}}, is 
\be
K_{ab}{}^i = - g_{\mu\nu} n^{\mu\, i} \nabla_a X_b^\nu
{= g_{\mu\nu} X_b^\nu \nabla_a n^{\mu i}}\,,
\ee
and $K^i = \gamma^{ab} K_{ab}{}^i $ denotes the mean extrinsic curvature. 
The twiddle on the covariant derivative of the normal vector in (\ref{eq:gw2}) denotes the $O(N-d)$ rotation covariant derivative
$\widetilde\nabla_a n^\mu{}_i = \nabla_a n^\mu{}_i - \omega_a{}_i{}^j n^\mu{}_j $, where the connection $\omega_a{}^{ij} = - \, \omega_a{}^{ji}$ is the
{\textit{twist potential}}, see {\it e.g.} \cite{defos}.
In this brief introduction, we prefer a `kinematical' description, without getting into the more essential
aspects of the embedding geometry, that involve a proper treatment of the relationship of the intrinsic and extrinsic geometry of the worldvolume,
via the Gauss-Codazzi-Mainardi equations.

\section{General covariant variational approach}\label{GeneralVar}

We consider a geometric model that describes the dynamics of a relativistic brane that satisfies the requirements of
reparametrization invariance and that is a scalar under background diffeomorphism transformations. In addition, we focus 
our attention on models that depend at most on first derivatives of the embedding functions
\be
S [X] = \int_w {\mathcal L} ( X_a^\mu)\,,
\label{eq:lagra}
\ee
where the Lagrangian density of weight one ${\mathcal L} (X^\mu_a)$ depends only on the intrinsic geometry of the worldvolume $w$. We have absorbed the differential $d^{d+1} \xi$ in the 
integral sign, henceforth. For the sake of simplicity we assume that the brane has no boundary,  and with appropriate  boundary conditions
 $\partial w = 0$. The simplest example that will be the focus of our attention in the next section is the DNG model, proportional to the volume of $w$,  with $\mu$ a constant tension,
\be
S_{DNG} [X]  = - \mu \int_w \sqrt{-\gamma}\,.
\label{eq:dng}
\ee
In the following, we consider the first, second, and third variations of the general intrinsic geometric model (\ref{eq:lagra}). This general approach allows
to establish a covariant perturbation theory that can be extended to more complex physical systems, once a proper framework for 
geometric models is introduced.

\subsection{First variation}

We consider an infinitesimal  variation of the field variables, the embedding functions, 
\be
X^\mu \to X^\mu + 
  {s}\delta X^\mu\,.
  \ee
 The standard field variation in the calculus of variations involves considering a one-parameter family of embedding functions  $X^\mu (\xi^a , s )$, with $s$ an arbitrary parameter that labels
  the leaves of the family, and a coordinate field variation $\delta X^\mu = (\partial X^\mu (\xi^a ,s ) / \partial s)|_{s=0}$, see {\it e.g.} \cite{GF,Kot}. 
 This `coordinate Taylor approach' is then  extended to higher orders, in  a systematic perturbation  treatment. Although appropriate and sufficient for systems with a finite number of degrees of freedom,
  as in  familiar classical mechanics systems,  in field theory it is convenient to upgrade the variation to take into account the underlying symmetries. 
  Thus, for our purposes, we adopt  instead a covariant directional variational derivative 
  \be
  \delta_X =  \delta X^\mu \nabla_\mu \,,
  \ee
   along the variation $\delta X^\mu$. For a treatment of the covariant variational derivative see {\it e.g.} \cite{bz77b},\cite{defos}, but we have not been able to identify  other
 more complete studies on this particular subject  in the available literature.  The manifest convenience  is that $\delta_X g_{\mu\nu} = 0$, since the background covariant derivative is assumed to be metric compatible. Moreover, we assume from the outset that the variation is conserved along the worldvolume.
Geometrically, this translates into the statement that the Lie derivative of the variation vector along the tangent vectors vanishes,
\be
[ \delta_X , \nabla_a ]X^\mu =  \delta_X X^\nu_a - \nabla_a \delta X^\nu = 0\,,
\label{eq:lie}
\ee
 or, more prosaically, that variation and worldvolume partial derivative commute. 
 {This can be understood from the fact that we are considering a one-parameter family
of embedding functions, labeled by $s$,  and infinitesimally partial derivatives of the embedding functions commute, as required
to have a foliation} \cite{GF}. It should also be emphasized that we resist the temptation to split the variation $\delta X^\mu$ into its normal and tangential parts. Although this is extremely convenient
 in the first variation, since the tangential variation can be associated to a mere reparametrization, and in the absence  
 of boundaries it can be safely neglected as pure gauge, at higher orders this early gauge fixing meddles the general structure of the perturbation theory.
 The usefulness of a spacetime covariant treatment in brane mechanics has been championed especially by Carter, see {\it e.g.} \cite{Carterbrane} and references therein. 
 
Turning to the first variation of the geometric model (\ref{eq:lagra}), we have simply
\be
\delta_X S[ X ] = \int_w \delta_X {\mathcal L} (X^\mu_a )\,,
\label{eq:delta}
\ee  
where we use (\ref{eq:lie}) to obtain
\be
\delta_X {\mathcal L} = { \partial {\mathcal L} \over \partial X^\mu_a } \; \delta X^\mu_a 
=  { \partial {\mathcal L} \over \partial X^\mu_a } \; \nabla_a \delta X^\mu\,.
\label{eq:deltaL}
\ee
Integration by parts of (\ref{eq:delta}) gives immediately
\ba
\delta_X S[ X ] &=& \int_w  { \partial {\mathcal L} \over \partial X^\mu_a } \; \nabla_a \delta X^\mu \nn \\
&=& \int_w {\mathcal E}_\mu (\mathcal{L}) \; \delta X^\mu + \int_w \nabla_a {\mathcal Q}^a\,,
\label{eq:first}
\ea  
where we identify the Noether current density 
\be
{\mathcal Q^a} = {\partial {\mathcal L} \over \partial X^\mu_a } \delta X^\mu
= {\mathcal P}_\mu{}^a \, \delta X^\mu 
\,,
\ee
 and  the canonical linear momentum, a spacetime 1-form and a worldvolume vector density,  
\be
{\mathcal P}_\mu{}^a =
{\partial {\mathcal L} \over  \partial X^\mu_a }\,.
\label{eq:mom}
\ee
The  Euler-Lagrange derivative is
\be
\begin{split}
 {\mathcal E}_\mu (\mathcal{L})&= -  \nabla_a \left(  {\partial {\mathcal L} \over \partial X^\mu_a } \right) = - \nabla_a {\mathcal P}_\mu{}^a \,.
 \end{split}
\ee
As a consequence of reparametrization invariance, it can be expressed as a conservation law.
The vanishing of the first variation (\ref{eq:first}), for arbitrary variations $\delta X^\mu$, produces the equations of motion 
\be
{\mathcal E}_\mu (\mathcal{L})= 0\,.
\label{eq:eom}
\ee
For intrinsic geometric models these equations of motion are of second order. This feature is shared by Lovelock branes, see {\it e.g.} \cite{CruzE}. For models that depend on the extrinsic geometry, for example on the squared extrinsic curvature, the background spacetime curvature appears already in the equations of motion \cite{Carterbrane,defos}. Higher derivative
geometric models can be studied in a covariant framework, but the reader should be warned that the treatment becomes quite complicated from the outset, even in a flat background spacetime. Yet, all the tools are available, and it can be carried
through, see {\it e.g.} \cite{Lovelock}.

\subsection{Second variation}

Once one has calculated the first variation, the second variation is the natural next step and it plays a relevant role in many physical contexts. For example, when one is studying structure formation in our Universe, or anisotropies in the Cosmic Microwave Background. In brane scenarios, the second variation is also important to study the stability of branes and know the behavior of their intrinsic geometry when a deformation is done. Besides, in a possible quantization via path integral of this kind of geometrical models, the second variation will be important to calculate the quantum correction to one loop.  \\

For the second variation of the geometric model (\ref{eq:lagra}), we choose to exploit the expression (\ref{eq:first}) for the first variation.
For the sake of symplicity, we assume that the worldvolume has no boundary, $\partial w = 0$, or that appropriate boundary conditions have been chosen, and we neglect the Noether current contribution for the moment. 

In the standard calculus of variations, one considers a one parameter family of embeddings $X^\mu (\xi^a; s)$ and a coordinate variation
\be 
\delta^2 X^\mu = 
\left( {\partial^2 X^\mu (\xi^a; s) \over  \partial s^2 } \right)_{s=0}\,.
\label{eq:csec}
\ee
As in the case of the first variation, we adopt a second order covariant variational derivative,
with appropriate labels, $\delta^2_{X} = \delta X^\mu \nabla_\mu(  \delta X^\nu \nabla_\nu \ ) $.
Then,  the second variation can be written as
\be
\delta^2_{X} S [X] = \int_w \left[  \left( \delta_{X} {\mathcal E}_\mu (\mathcal{L}) \right) \; \delta X^\mu + {\mathcal E}_\mu (\mathcal{L}) \; \delta^2 X^\mu  \right]\,.
\ee
Assuming that the equations of motion (\ref{eq:eom}) are satisfied  the second term vanishes, and this expression reduces to
\be
		\delta^2_{X}S [X] 
			\big |_{[0]}
		= \int_w \left[ \delta_{X} {\mathcal E}_\mu (\mathcal{L})  \big |_{[0]} \right] \; \delta X^\mu \,,
\label{eq:sec1}
\ee
		where the subscript 
			$[0]$
		is a reminder that we are on-shell.
The evaluation of the second variation amounts to a  linearization of the equations of motion about a solution, 
{ {and this is the path chosen by various authors, with a direct evaluation of $ \delta_{X} {\mathcal E}_\mu (\mathcal{L})  \big |_{[0]}$  \cite{guven1993covariant, garriga1991perturbations, LF94}.
In this note, we explore a different path, taking full advantage of the conservation law structure of the equations of motion.}}
It should be emphasized that, in contrast to \eqref{eq:lie}, when applying a variation to a worldvolume tensorial quantity it no longer 
commutes with a background covariant derivative. This is a  consequence of the fact that the variation operator
is a directional covariant derivative. Indeed, this is where the background spacetime curvature
comes into play, as shown explicitly in the following.
We have
\be
\begin{split}
\delta_{X} {\mathcal E}_\mu (\mathcal{L})  
			\big |_{[0]}
&= - \delta_{X} \nabla_a {\mathcal  P}_\mu{}^a\\
& = -  [ \delta_{X} , \nabla_a ] {\mathcal P}_\mu{}^a - 
\nabla_a \delta_{X} {\mathcal P}_\mu{}^a \\
 &= - R_{\alpha\beta\mu}{}^\rho \delta X^\alpha X^\beta_a {\mathcal P}_{\rho}{}^a -  \nabla_a \delta_{X} {\mathcal P}_\mu{}^a\, ,
 \end{split}
\ee
where we have used the Bianchi identity {in the first term, that acts} on a spacetime 1-form $\omega_\rho${ as} $ [\nabla_\mu, \nabla_\nu] \omega_\rho = R_{\mu\nu \rho}{}^\sigma \omega_\sigma${.}  
Therefore the  second variation (\ref{eq:sec1})  takes the form
\be
 \delta^2_{X}  S [X] \big |_{[0]} = -  \int_w \left[ \nabla_a \delta_{X} {\mathcal P}_\mu{}^a + R_{\alpha\beta\mu}{}^\rho \delta X^\alpha X^\beta_a {\mathcal P}_{\rho}{}^a \right] \delta X^\mu{} \,.
  \label{eq:sec2}
  \ee
  Using the definition of the canonical momentum (\ref{eq:mom}), the first term can be written as its variation, or linearization,
 \be
 \begin{split}
 \delta_{X}  {\cal P}_\mu{}^a  &= {\delta^2 {\cal L} \over \partial X^\nu_b \partial X^\mu_a} (\delta_{X} X^\nu_b ) \\
 &= {\cal H}_{\nu\mu}^{ba} (\delta_{X} X^\nu_b )  \\
 &= {\cal H}_{\nu\mu}^{ba} (\nabla_b \delta X^\nu ) \,,
 \label{eq:hess}
 \end{split}
\ee
where we have defined the Hessian matrix
 \be
  {\cal H}_{\nu\mu}^{ba}  = {\delta^2 {\cal L} \over \partial X^\nu_b \partial X^\mu_a}\,,
  \label{eq:hess1}
 \ee
 and in the last equality we have used the fact that variation and partial derivative commute, see (\ref{eq:lie}).
 Note that the Hessian is degenerate, it admits null eigenvectors, because of the gauge freedom associated with reparametrization invariance.
 The Hessian is symmetric in mixed pairs of indices
 \be
  {\cal H}_{\nu\mu}^{ba} =  {\cal H}_{\mu\nu}^{ab}\,.
  \label{eq:symm}
  \ee

Inserting (\ref{eq:hess}) in the  second variation (\ref{eq:sec2}),  one obtains the general expression 
\be 
\begin{split}
\delta^2_{X}  S [X] \big |_{[0]} =- \int_w \left\lbrace\right. &   \nabla_a  \left[ {\cal H}_{\nu\mu}^{ba} (\nabla_b \delta X^\nu ) \right] \\
& +  \left.  R_{\alpha\beta\mu}{}^\rho \delta X^\alpha X^\beta_a {\mathcal P}_{\rho}{}^a \right\} \delta X^\mu\,.
 \label{eq:sec3}
 \end{split}
  \ee

All that is left to do is the calculation of the Hessian matrix, for a given  geometric model defined by a specific
${\mathcal L}$. 

{At this point, $\delta X^\mu$ is arbitrary, and represents a linearized perturbation around any on-shell embedding. However, if we are interested in perturbations from one on-shell configuration to another, we need to choose $\delta X^\mu$ so that the second perturbation vanishes. From }(\ref{eq:sec3}){ we can immediately read the necessary non trivial set of equations for the vector $\delta X^\mu$. These set of equations are known as \textit{Jacobi equations}.}
We set the vanishing of the second variation in the form 
\be 
\delta^2_{X} S [X] \big |_{[0]} = 0\,, 
\label{eq:jac0}
\ee
for {the} variation $ \delta X^\mu$, that we {can} identify as  a worldvolume `deviation vector', $ \delta X^\mu = \eta^\mu$, { }interpreted as a {linear approximation} vector connecting neighboring  branes
in their evolution, just like in the familiar case of the geodesic deviation equation (see e.g. \cite{Wald}, \cite{bz89,Bazpol}). 
{Expression }(\ref{eq:jac0}) gives the Jacobi equations for the deviation vector $\eta^\mu$  in the form
\be
\begin{split}
  - \nabla_a  \left[ {\cal H}_{\nu\mu}^{ba} \; \nabla_b  \eta^{\nu} \right]  
  - R_{\nu\beta\mu}{}^\rho  X^\beta_a {\mathcal P}_{\rho}{}^a \eta^{\nu}\,= 0\,,
\label{eq:jac1}
\end{split}
\ee
the first `kinetic' term is of second order, as expected, since the equations of motion are of second order. It displays the central role 
of the Hessian as a sort of `mass-matrix' in the language of classical mechanics. 
The second term, involving the background spacetime, can be seen as an external force. This is consistent with our 
intuition based on the geodesic deviation equation in General Relativity \cite{Wald}.
It is worth noticing that in the case of a flat background, the Jacobi equations also take the form of a conservation law, as a divergence free
equation for the linearized canonical momentum. For an alternative, and more economic, avenue to the brane Jacobi equations in a curved background spacetime, 
using a covariant simultaneous variational principle, see \cite{CG1}. 

{Going back to }(\ref{eq:sec3}){, another thing we can do is an integration by parts of the first term, which} produces, neglecting a boundary term, 
\be
\begin{split}
  \delta_{X}^2  S [X] \big |_{[0]} =  \int_w \left\{  \right. & {\cal H}_{\nu\mu}^{ba} (\nabla_b \delta X^\nu ) (\nabla_a \delta X^\mu )\\ 
  &\left.- R_{\alpha\beta\mu}{}^\rho \delta X^\alpha X^\beta_a {\mathcal P}_{\rho}{}^a \delta X^\mu \right\} \nn \\
  =
  \int_w \, {\cal I}& (\delta X , \delta X )\,.
  \label{eq:index}
 \end{split}
 \ee
This expression identifies the quadratic form, or index, ${\cal I} (\delta X , \delta X )$, { { that is of fundamental importance in establishing stability of the
solutions of the equations of motion}}, see {\it e.g.}  \cite{GF} in the calculus of variations, and \cite{FT} for minimal surfaces in Euclidean three-dimensional ambient space.
{The index may be evaluated at two different deviation vectors, ${\cal I} (\delta X_{(1)} , \delta X_{(2)} )$, corresponding with two different deviation directions.}
Usually the variations are identified, $ \delta X_{(2)} = \delta X_{(1)} = \eta $, and { {one of the main difficulties in a proper analysis of the quadratic form  lies in finding an appropriate complete basis
that satisfies the required  boundary conditions, depending on the problem at hand, to allow for a proper understanding of the quadratic form}}  \cite{GF,GH}.
This idea  of using different deviation vectors, and explore how it affects the stability analysis, will be considered in  future work.
The expression (\ref{eq:index}) is, except for factors of one half, identical to the Jacobi accessory variational principle, that 
provides an action whose extremalization  gives 
the Jacobi equation as the vanishing of its Euler-Lagrange derivative, see {\it e.g.} \cite{Kot}.

\subsection{Third variation}

{ { In order to establish the usefulness of a covariant variational approach to perturbation theory for the dynamics of branes,
we  extend our treatment to third order.  In addition, a reason to recur to the third variation of the geometric model 
is a possible degeneracy of the contribution at second order between deformation modes. }}
As expected, the complexity of the calculation increases noticeably, yet a  covariant variational approach allows to appreciate the underlying mathematical
variational 
structure, and makes it ready for use in concrete
applications when second order perturbations are included. 
Moreover, the covariant  approach clarifies the 
interplay between first order and second order perturbations. For a previous study of these issues, see \cite{Edgarthesis}.

We consider the expression (\ref{eq:sec1}) for the second variation of the action. An additional variation (corresponding to an additional derivative in (\ref{eq:csec})), gives
\be
\begin{split}
\delta_{X}^3 S [X] \big |_{[0]} = \int_w &\left\lbrace  \left[ \delta_{X}^2 {\mathcal E}_\mu (\mathcal{L})  \big |_{[0]} \right]  \right.\; \delta X^\mu \\
&\left.+ \left[ \delta_{X} {\mathcal E}_\mu (\mathcal{L})\big  |_{[0]}  \right] \;  \delta^2 X^\mu  \right\}
 \, .
\label{eq:sec31}
\end{split}
\ee
Assuming that both the equations of motion (\ref{eq:eom}) and the Jacobi equations (\ref{eq:jac1}) are satisfied,
then the second term vanishes, and
the third variation (\ref{eq:sec31})   reduces to
\be
\delta_{X}^3 S [X] 
			%|_{(0,1)} 
			\big |_{[0,1]}
= \int_w  \left[ \delta_{X}^2 {\mathcal E}_\mu (\mathcal{L})  \big|_{[0,1]} \right] \; \delta X^\mu \,,
\label{eq:doubleshell}
\ee
where now the subscript $[0,1]$ is a reminder that, besides the equations of motion, 
also the Jacobi
equations  are assumed to be fullfilled.
It remains to unpack the second variation of the equations of motion. \\
For this purpose, it is 
convenient to first assume a flat background, with vanishing Riemann tensor, and then in a second step  introduce
the complication of a curved background.

\subsubsection{Flat background spacetime}

Within the special case of a Minkowski spacetime,  background derivatives commute, and we exploit, just like for the second variation, 
the fact that the equations of motion can be written
as a conservation law.

In terms of the linear momentum, the second variation is given by (\ref{eq:sec2}). 
Therefore, using  (\ref{eq:mom}) , the third variation is given by
\be
\delta_X^3 S [ X ] \big|_{[0,1]} =   
\int_w  \left[ \left( \nabla_a \delta^2 {\mathcal P}_\mu{}^a \right)    \delta X^\mu \right] \big|_{[0,1]},
\label{eq:third}
\ee
since partial derivative and variation commute.
To unpack the second variation of the linear momentum, we use (\ref{eq:hess}) for  the first variation of the linear momentum, and remark that the variation,
when acting on the first variation of the shape functions  $\delta X$ gives a second variation of the shape functions $\delta^2 X$, and when acting on the Hessian matrices
of the energy density gives a source term that depends quadratically on the first variations $\delta X$. It is important to remember that
the assumption that the Jacobi equation is satisfied is equivalent to assume that the first order perturbations  $\delta X$ are given. In this sense, 
we are talking of a source term.
The second variation of the linear stress tensor, as given by the variation of (\ref{eq:hess}),  can then be written as
the sum
\be
\delta_X^2 {\mathcal P}_\mu{}^a (\delta^2 X, \delta X ) = \delta_X {\mathcal P}_\mu{}^a (\delta^2 X ) + {\mathcal S}_\mu{}^a (\delta X).
\label{eq:d2f}
\ee
The first term depends on the second variation  $\delta^2 X$, and  it has the same structure of the first order variation.
It is given by the Jacobi operator, with $\delta X \to \delta^2 X$ in (\ref{eq:hess}). 
 This term does not require any additional work, since the Jacobi operator is known, see (\ref{eq:jac1}).
The second term in (\ref{eq:d2f})  is a source term that depends on the first variation $\delta X$, due to the variation
of the Hessian matrix, 

\be
 {\mathcal S}_\mu{}^a (\delta X) = \left( {\partial^2  \delta {\mathcal L} \over \partial X^\mu_a \partial X^\nu_c } \right) \nabla_c \delta X^\nu\,.
\ee
This term does require additional work.
Using (\ref{eq:deltaL}) for $\delta {\mathcal L}$, it takes the formidable aspect
\be
{\mathcal S}_\mu{}^a (\delta X)   =    {\cal T}^{cba}_{\rho\nu\mu}  \;  \nabla_c \delta X^\nu 
\nabla_b \delta X^\rho\, ,
\ee
where we have defined the tensor of third order derivatives analogous to the Hessian
\be
{\cal T}^{cba}_{\rho\nu\mu}={\partial^3 {\cal L} \over \partial X^\rho_c \partial X^\nu_b \partial X^\mu_a}.
\label{eq:T3}
\ee
Returning to the third variation of the action (\ref{eq:third}),  we find therefore that it can be written in the
deceivingly simple form
\be
\delta^3 S [ X ] \big|_{[0,1]} = \int_w \left\{ \left[ \mathcal{J}_\mu (\delta^2 X)  + \nabla_a {\mathcal S}_\mu{}^a (\delta X) \right]
 { \delta X}^\mu \right\} \big|_{[0,1]},
\label{eq:terza}
\ee
where the second order perturbations appear in the Jacobi operator, $ {\mathcal J}_\mu (\delta^2 X) =
\nabla_a  \delta {\mathcal P}_\mu{}^a (\delta^2 X)$, and the first order perturbations in the source term $\nabla_a 
{\mathcal S}_\mu{}^a (\delta X) $.
 What can be appreciated from this expression for the third variation is that the higher order 
Jacobi equation it implies  has a very different structure than the second order Jacobi equation.
In particular, it is not homogenous, for the presence of a source term.
Also for the third variation one can derive an accessory variational principle and a parallel simultaneous variational 
principle.

\subsubsection{Curved background spacetime}

Let us now consider how the third variation (\ref{eq:terza}) is modified when the brane evolves in a curved background, when partial derivatives and variations do not commute.
We start with equation (\ref{eq:sec3}) for the second variation with a non-vanishing curvature term. An additional third variation yields
\be
\begin{split}
\delta_X^3 S\left[ X\right]\big\vert_{[0,1]}=-\int_{\omega}&\left\lbrace \nabla_a\left[\delta_X\left(\mathcal{H}_{\nu\mu}^{ba}\nabla_b\delta X^{\nu}\right)\right]\right.\\
&+R_{\alpha\beta\nu}{}^{\rho}\delta X^{\alpha}X^{\beta}_{a}\delta_X\mathcal{P}_{\rho}{}^{a}\\
&\delta_{X}\left( R_{\alpha\beta\mu}{}^\rho \delta X^\alpha X^\beta_a {\mathcal P}_{\rho}{}^a \right)
 \Big\} \delta X^\mu\, ,
 \label{eq:3vbulk}
\end{split}
\ee
where we used the Bianchi identity yet again {in the first line}. The first term can be written in terms of the canonical momentum using \eqref{eq:hess}
\begin{equation}
\begin{split}
\delta_{X}  \left (\mathcal{H}_{\nu\mu}^{ba} \nabla_b \delta X^\nu \right)=&\delta_{X}^2\mathcal{P}_\mu{}^a  \\
=&  \mathcal{T}^{cba}_{\rho\nu\mu} (\nabla_c \delta X^\rho)(\nabla_b \delta X^\nu)\\
&+  \mathcal{H}^{ba}_{\nu\mu}  \delta_{X}(\nabla_b \delta X^\nu)  \\
=& \mathcal{T}^{cba}_{\rho\nu\mu}(\nabla_c \delta X^\rho)(\nabla_b \delta X^\nu)\\
 &+\mathcal{H}^{ba}_{\nu\mu} R^\nu{}_{\alpha\rho\beta}\delta X^{\alpha}\delta X^{\rho}X^\beta_b\\
 &+ \mathcal{H}^{ba}_{\nu\mu} \nabla_b \delta^2 X^\nu,
  \label{eq:2vP}
 \end{split}
\end{equation}
where the tensor $\mathcal{T}$ is defined in (\ref{eq:T3}). 
In the last step, we commuted the derivative and the variation producing a background Riemann tensor projection. 
{Let us now use the Leibniz rule to unpack the third term in }\eqref{eq:3vbulk},
\begin{equation}
\begin{split}
\delta_{X}\left( R_{\alpha\beta\mu}{}^\rho \delta X^\alpha X^\beta_a {\mathcal P}_{\rho}{}^a\right)
=&\left( \nabla_\sigma R_{\alpha\beta\mu}{}^\rho{} \right) \delta X^{\sigma}  \delta X^\alpha X^\beta_a {\mathcal P}_{\rho}{}^a \\
&+R_{\alpha\beta\mu}{}^\rho {\left(\delta^2 X^\alpha\right)} X^\beta_a {\mathcal P}_{\rho}{}^a\\
&+ R_{\alpha\beta\mu}{}^\rho \delta X^\alpha {\nabla_a \left(\delta X^\beta\right)} {\mathcal P}_{\rho}{}^a\\
&+ R_{\alpha\beta\mu}{}^\rho \delta X^\alpha X^\beta_a \left(\delta_{X}{\mathcal P}_{\rho}{}^a\right),
\label{eq:3vR}
\end{split}
\end{equation}
{where we used }\eqref{eq:lie}{ in the second term.} Notice that we now have many curvature terms along with a derivative of the Riemann tensor which severely complicates the expression, compared to the case of flat background spacetime.
Inserting \eqref{eq:3vR} and \eqref{eq:2vP} into \eqref{eq:3vbulk}, we get
\begin{equation}
\begin{split}
 \delta_{X}^3  S [X] \big |_{[0,1]}= 
  &-  \int_w \Big\{ \nabla_a  \left[  {\cal T}^{cba}_{\rho\nu\mu}(\nabla_c \delta X^\rho)(\nabla_b \delta X^\nu)\right.\\
&\left. + {\cal H}^{ba}_{\nu\mu}\left( \nabla_b \delta^2 X^\nu+R^\nu{}_{\alpha\rho\beta}\delta X^{\alpha}\delta X^{\rho}X^\beta_b \right) \right]	 \\
 &+2R_{\alpha\beta\mu}{}^{\rho}X^{\beta}_a \delta X^{\alpha}  \delta_{X}{\cal P}^a_{\rho}\\ 
 &+\left( \nabla_\sigma R_{\alpha\beta\mu}{}^\rho{} \right) \delta X^{\sigma}  \delta X^\alpha X^\beta_a {\mathcal P}_{\rho}{}^a	 \\
&+R_{\alpha\beta\mu}{}^\rho {\left(\delta^2 X^\alpha\right)} X^\beta_a {\mathcal P}_{\rho}{}^a\\
&+ R_{\alpha\beta\mu}{}^\rho \delta X^\alpha {\left(\nabla_a\delta X^\beta\right)} {\mathcal P}_{\rho}{}^a
 \Big\} \delta X^\mu\,.
\end{split}
\label{eq:3vFull} 
\end{equation}
{As we can see, the first two lines are again a conservation law, now for the second variation of the canonical momentum. Additionally, we have a term proportional to both the curvature and the first variation of the momentum, and a term linear in the second order deviation vector. Notice that now we have quadratic terms and derivatives of the first order deviation vector. It is assumed however, that such vector is known.}\\
The vanishing of the previous integral leads to the equation
\begin{equation}
\begin{split}
&\nabla_a  \left[ \delta_{X}^2 \mathcal{P}_\mu{}^a \right] 
+ R_{\alpha\beta\mu}{}^{\rho}\delta X^{\alpha} X^{\beta}_a \delta_{X}\mathcal{P}_{\rho}{}^a\\
&+ \delta_{X}\left( R_{\alpha\beta\mu}{}^\rho \delta X^\alpha X^\beta_a {\mathcal P}_{\rho}{}^a\right)
=0,
\label{eq:3veom}
\end{split}
\end{equation}
where we conveniently wrote it in terms of the canonical momentum, both to avoid unnecessarily long expressions and to show, as promised, the underlying structure of the equations. Here, the embeddings are known from the equations of motion, and the variation vector $\delta X^\nu$ 
satisfies the Jacobi equation \eqref{eq:jac1} with $\eta^\nu = \delta X^\nu$. Equation \eqref{eq:3veom} now determines the second order variation vector {$\delta^2 X^\nu$}.\\

Like in the previous section, the variation \eqref{eq:3vFull} can be organized as follows
\begin{equation}
\begin{split}
\delta_X^3S[X]\big |_{[0,1]}=-\int_{\omega}\left[ \mathcal{J}_{\mu}(\delta^2 X)+\mathcal{F}_{\mu}(\delta X) \right]\delta X^{\mu},
\label{ethirdvar}
\end{split}
\end{equation}
where $\mathcal{F}_{\mu}$ is the source term given by 
\begin{widetext}
\begin{equation}
\begin{split}
\mathcal{F}_{\mu}(\delta X)=&\nabla_a\left[ \mathcal{T}^{cba}_{\rho\nu\mu}\left( \nabla_c\delta X^{\rho}\right)\left(\nabla_b\delta X^{\nu}\right) +\mathcal{H}^{ba}_{\nu\mu}R^{\nu}{}_{\alpha\rho\beta} \delta X^{\alpha}\delta X^{\rho} X_b^{\beta}\right]
+\left( \nabla_\sigma R_{\alpha\beta\mu}{}^\rho{} \right) \delta X^{\sigma}  \delta X^\alpha X^\beta_a {\mathcal P}_{\rho}{}^a \\
&+R_{\alpha\beta\mu}{}^{\rho}\left[ 2\delta X^{\alpha}X^{\beta}_{a}\mathcal{H}^{ba}_{\nu\rho}\nabla_b\delta X^{\nu}+\delta X^{\alpha}\nabla_a\delta X^{\beta}\mathcal{P}_{\rho}{}^a\right].
\label{sourceterm}
\end{split}
\end{equation}
\end{widetext}
Then, we can see that \eqref{eq:3veom} is nothing but a Jacobi equation with source $\mathcal{F}_\mu$ for the second variation field $\delta^2 X^\mu$, just like in \eqref{eq:terza}. However, here the source term  is modified by the curvature of the spacetime background. \\
The expression \eqref{eq:3veom} has already been found in \cite{KN14}, but there only normal perturbation modes were considered, and the perturbations were done directly in the equations of motion. Here a fully covariant expression is given and seen from a variational perspective. This expression is a second order approximation to the deviation of neighboring on-shell branes in a fixed arbitrary background. Going to second order perturbations is not only a natural next step after first order. There are other reasons to recur to second order perturbations such as overcoming degeneracies that appear at first order. Some physical quantities have a second order leading contribution. For a more extensive list, see {\it e.g.} \cite{Mars}.

\section{Dirac-Nambu-Goto brane} \label{DNGbrane}
The  DNG action is a first approximation in this approach to constructing geometric models to describe physical systems,
where we put to use the general covariant variational formulation we have described above.
Originally, it was proposed by Dirac in a geometric relativistic model for an `extensible electron' \cite{dirac1962extensible}. Later, Nambu and independently Goto applied the same idea to a relativistic string to model gluons in QCD \cite{nambu1970duality, goto1971relativistic}. The rest is a long history.\\
From the viewpoint of brane mechanics, the DNG action is an obvious starting point.
This action is proportional to the  worldvolume of the brane and  is invariant under diffeomorphisms. The DNG action is given by
\be
\begin{split}
S_{DNG} [ X] &= \int_w {\mathcal L}_{DNG} =  - \mu \int_w \sqrt{-\gamma}\, ,
\end{split}
\ee
 where $\mu$ is the membrane tension. Using the framework previously developed, for the first variation we only need the canonical momentum. The canonical momentum is
\be
\begin{split}
{\cal P}_\mu{}^a &= {\partial {\cal L}_{DNG} \over  \partial X_a^\mu  } =  - \, \mu \sqrt{-\gamma} \, g_{\mu\nu} \gamma^{ab} X^\nu_b \,.
\label{eq:DNGP}
\end{split}
\ee
Note that it is tangential to the worldvolume, as is the case for any intrinsic geometric model and isotropic.
The equations of motion that are given by the vanishing of the first variation are
\be
\begin{split}
{\mathcal E}_\mu ({\mathcal L}_{DNG}) & = -\nabla_a {\mathcal P}_\mu{}^a = -\mu \nabla_a \left( \sqrt{-\gamma} \, g_{\mu\nu} \gamma^{ab} X^\nu_b \right)=0.
\end{split}
\ee
Since the action is invariant under reparametrizations, the tangential variations do not influence the dynamics because these can be identified as a divergence. Thus when one projects the equations of motion onto the basis of the background spacetime, only the normal part survives, the latter is given by  
\be
n^\mu{}_i \nabla_a {\mathcal P}_\mu{}^a  = \mu \sqrt{-\gamma} K_i
 = 0\,,
\label{NullK}
\ee
{\it i.e.} the vanishing of the worldvolume mean extrinsic curvature $K_i = \gamma^{ab} K_{ab\,i}$.
 This is the relativistic version of the equilibrium condition for a minimal surface with arbitrary codimension. \\

For the second variation, we need only to calculate the Hessian,
\be
\begin{split}
 {\cal H}_{\nu\mu}^{ba}  = {\delta^2 {\mathcal L_{DNG}} \over \partial X^\nu_b \partial X^\mu_a}=   - \mu \sqrt{-\gamma}   \left(
 \gamma^{ab} n_{\mu\nu} + X_{\mu\nu}^{ab} \right) \,,
 \label{eq:DNGH}
 \end{split}
\ee
where we have used the completeness relationship (\ref{eq:comple}) 
that  defines the  normal projector $n^{\mu\nu}$. 
The tangential bivector that appears in (\ref{eq:DNGH}) is 
\be
X^{\mu\nu}_{ab} = 2X^{[\mu}_a X^{\nu]}_b \,,
 \ee 
with indices raised with $\gamma^{ab}$ and  lowered with $g_{\mu\nu}$. 
Substituting in the general expression for the second variation \eqref{eq:sec3} we have
\be
\begin{split}
\delta_{X}^2 S[X]\big\vert_{[0]}
=\mu \int_w  &
\left\{ \nabla_a\left[ \sqrt{-\gamma}(\gamma^{ab}n_{\mu\nu}+X^{ab}_{\mu\nu})(\nabla_b \delta X^\nu)\right]\right.\\
&+\left.\sqrt{-\gamma}R_{\alpha\beta\mu\rho}\delta X^\alpha h^{\beta \rho}\right\} \delta X^{\mu},
\end{split}
\ee
where we identified the tangential projector defined in the completeness relation \eqref{eq:comple}, given by
\be
h^{\beta\rho}=\gamma^{bc}X^\beta_b X^\rho_c.
\ee
For this case the Jacobi equations \eqref{eq:jac1} read 
\be
\begin{split}
&\mu \left\{ \nabla_a \left[\sqrt{-\gamma}  \left(\gamma^{ab} n_{\mu\nu} + X_{\mu\nu}^{ab} \right) (\nabla_b \delta X^\nu ) \right]\right.\\
  &+ \left.\sqrt{-\gamma} R_{\alpha\beta\mu\rho} \delta X^\alpha h^{\beta\rho}\right\}=0\, .
  \label{secondDNG}
 \end{split}
\ee
Notice that normal and tangential projections of the deviation vector arise naturally in the expression, showing the mixture of both contributions. These equations provide a generalization to branes of the well known geodesic deviation equation for particles.\\

For the third variation we need to obtain the tensor $\cal T$, defined in (\ref{eq:T3}). When the algebraic dust settles down, we obtain
\be
\begin{split}
{\cal T}^{cba}_{\tau\nu\mu}=&\sqrt{-\gamma}\left\{ n_{\nu\tau}(\gamma^{bc}X^a_\mu-2\gamma^{ab}X^c_\mu) \right.\\
&+\left. 2n_{\mu\tau}(\gamma^{ac}X^b_\nu-\gamma^{ab}X^c_\nu-\gamma^{bc}X^a_\nu) + X^{abc}_{\mu\nu\tau}	\right\}\, ,
\label{eq:DNGT}
\end{split}
\ee
where there is an abundance of raised and lowered indices, and now we have a tangential antisymmetric trivector
\be
X^{abc}_{\mu\nu\tau}:=3!X^{[a}_{\mu}X^{b}_{\nu}X^{c]}_{\tau}\, .	
\ee
{Putting it all together, we use the expressions \eqref{eq:DNGP},\eqref{eq:DNGH}, and \eqref{eq:DNGT} { to insert them into }\eqref{ethirdvar}, taking into account \eqref{sourceterm}. Then, we can write the full third variation for the DNG action}
\begin{widetext}
\be
\begin{split}
\delta_X^3S_{DNG}\big |_{[0,1]}=&  \int_{\omega}\left\lbrace \mathcal{J}_{\mu}(\delta^2 X)+\mu\nabla_{a}\left[ \sqrt{-\gamma}\left( n_{\nu\tau}(\gamma^{bc}X^a_\mu-2\gamma^{ab}X^c_\mu) 
+ 2n_{\mu\tau}(\gamma^{ac}X^b_\nu-\gamma^{ab}X^c_\nu-\gamma^{bc}X^a_\nu) + X^{abc}_{\mu\nu\tau}	\right) \right.\right.\\
 &\left. \left(\nabla_c \delta X^{\tau}\right)\nabla_b\left( \delta X^{\nu}\right) +\sqrt{-\gamma}   \left(
 \gamma^{ab} n_{\mu\nu} + X_{\mu\nu}^{ab} \right)R^{\nu}{}_{\alpha\rho\beta}\delta X^{\alpha}\delta X^{\rho}X_b^{\beta}\right]+\mu\sqrt{-\gamma}\left(\nabla_{\omega}R_{\alpha\beta\mu\rho}\right)\delta X^{\omega}\delta X^{\alpha}h^{\beta\rho}\\
 &\left.+\mu \sqrt{-\gamma} R_{\alpha\beta\mu}{}^{\rho}\left[ 2\delta X^{\alpha}X_a^{\beta}\nabla_b(\delta X^{\nu}) \left( \gamma^{ab}n_{\rho\nu}+X^{ab}_{\rho\nu}\right)+\delta X^{\alpha}\nabla_a(\delta X^{\beta})g_{\rho\nu}\gamma^{ac}X_c^{\nu} \right]\right\rbrace\delta X^{\mu}.
\end{split}
\ee
\end{widetext}
Of course, the vanishing of the last integral gives us the second order perturbation to the equations of motion
\begin{widetext}
\begin{equation}
\begin{split}
 \mathcal{J}_{\mu}(\delta^2 X)
 =&-\mu\left\lbrace\nabla_{a}\left[ \sqrt{-\gamma}\left[\left( n_{\nu\tau}(\gamma^{bc}X^a_\mu-2\gamma^{ab}X^c_\mu) 
+ 2n_{\mu\tau}(\gamma^{ac}X^b_\nu-\gamma^{ab}X^c_\nu-\gamma^{bc}X^a_\nu) + X^{abc}_{\mu\nu\tau}	\right) \right. \left(\nabla_c \delta X^{\tau}\right)\nabla_b\left( \delta X^{\nu}\right)\right.\right.\\
&+\left.\left.  \left( \gamma^{ab} n_{\mu\nu} + X_{\mu\nu}^{ab} \right)R^{\nu}{}_{\alpha\rho\beta}\delta X^{\alpha}\delta X^{\rho}X_b^{\beta}\right]\right]+\sqrt{-\gamma}\left(\nabla_{\omega}R_{\alpha\beta\mu\rho}\right)\delta X^{\omega}\delta X^{\alpha}h^{\beta\rho}\left.\right.\\
&+\left. 
\sqrt{-\gamma} R_{\alpha\beta\mu}{}^{\rho}\left[ 2 \left( \gamma^{ab}n_{\rho\nu}+X^{ab}_{\rho\nu}\right)\delta X^{\alpha}X_a^{\beta}\nabla_b\delta X^{\nu}+g_{\rho\nu}\gamma^{ac}X_c^{\nu} \delta X^{\alpha}\nabla_a\delta X^{\beta}\right]\right\rbrace ,
\end{split}
\label{fullDNG3}
\end{equation}
\end{widetext}
here $\mathcal{J}_{\mu}(\delta^2 X)$ is given by the left side of \eqref{secondDNG}, with $\delta X\rightarrow \delta^2 X$. The right side of \eqref{fullDNG3} is the source term of Jacobi's nonhomogeneous equation. Notice that the source term is influenced by the Riemann tensor of the background spacetime and its covariant derivative. One might think that the covariant derivative on the Riemann tensor should vanish since the background spacetime is fixed. However, when making variations, local changes in the embeddings must be taken into account, which justifies the appearance of this term.

\section{Discussion}\label{Discussion}

In this paper we have developed a systematic covariant variational approach to perturbation theory for intrinsic brane models.
Perturbations are realized as covariant derivatives along the flow of a deformation vector, with respect to a one parameter family of the
worldvolume corresponding to the brane's evolution. These vectors came from
coordinate variations of a two parameters family of embeddings $X^{\mu}(\xi^a; s)$. This is the standard approach used in the calculus of variations, but it has to be modified appropriately for systems with an infinite number of degrees of freedom.

As a first step,
we consider models that depend only on the intrinsic geometry, {\it i.e.} that contain only  first derivatives of the embedding functions.
This set-up allows for a generalization to higher derivatives, relevant in many contexts, with the obvious complications involved.

We introduce a general covariant variational approach that to first order reproduces well known equations of motion.
At second order, we obtain the Jacobi equations that describe the evolution of deviation vectors that depict the behavior of a  one-parameter
family of branes under the external force provided by the background projected Riemann curvature. In the kinetic term that appears
in the Jacobi equations it is emphasized the central role of the Hessian matrix of the geometrical model. In the special case of a DNG geometric model, the kinetic term takes the form of a `mass matrix' that separates neatly in a normal and tangential contribution.
It should be mentioned at this point that a fully covariant approach is needed especially when considering boundary contributions.
This consideration is relevant when issues of boundary vs. bulk contributions are deemed to be relevant.
The Jacobi equation itself
can be obtained from an accessory variational principle, as its Euler-Lagrange equation.\\
In our search for a suitable geometric structure
for higher order variations, we extend the approach to a third variation of the action, making contact with previous investigations.
What emerges is that the structure of the third variation is qualitatively different. The kinetic term is related to the Jacobi operator that
appears in the second variation, but there is also a source term. This indicates the existence of a generic structure in perturbation
theory that needs to be explored in more detail.

The equations obtained are very general, and can be applied readily  to any specific background and any geometric model of interest.
Unlike previous related works, we have not separated the perturbation into tangential and normal modes. Although this means we have non-physical degrees of freedom in our results, one can always substitute normal deformations to eliminate this, recovering the results of previous works.
The variational approach allowed us to notice that all perturbations may be expressed as conservation laws plus curvature terms.

The Dirac-Nambu-Goto action was taken as an example. We found that beyond first order, the tangential and normal projections of the deformation vectors are naturally present,
even though the vector itself remained covariant. Even in this simpler case, the second order perturbation equations can be very complicated and the contribution of the spacetime curvature is highly relevant.
The framework developed here can be applied to any intrinsic model that depends at most on first order derivatives of the embedding functions,
the case for higher order dependance may be developed in a similar way.

\section*{Acknowledgements}
GC and EYL thank CONACyT for their graduate fellowships. RC thanks Sistema Nacional de Investigadores for partial support.
\bibliography{HigherOrderPerts}

\begin{thebibliography}{42}%
\makeatletter
\providecommand \@ifxundefined [1]{%
 \@ifx{#1\undefined}
}%
\providecommand \@ifnum [1]{%
 \ifnum #1\expandafter \@firstoftwo
 \else \expandafter \@secondoftwo
 \fi
}%
\providecommand \@ifx [1]{%
 \ifx #1\expandafter \@firstoftwo
 \else \expandafter \@secondoftwo
 \fi
}%
\providecommand \natexlab [1]{#1}%
\providecommand \enquote  [1]{``#1''}%
\providecommand \bibnamefont  [1]{#1}%
\providecommand \bibfnamefont [1]{#1}%
\providecommand \citenamefont [1]{#1}%
\providecommand \href@noop [0]{\@secondoftwo}%
\providecommand \href [0]{\begingroup \@sanitize@url \@href}%
\providecommand \@href[1]{\@@startlink{#1}\@@href}%
\providecommand \@@href[1]{\endgroup#1\@@endlink}%
\providecommand \@sanitize@url [0]{\catcode `\\12\catcode `\$12\catcode
  `\&12\catcode `\#12\catcode `\^12\catcode `\_12\catcode `\%12\relax}%
\providecommand \@@startlink[1]{}%
\providecommand \@@endlink[0]{}%
\providecommand \url  [0]{\begingroup\@sanitize@url \@url }%
\providecommand \@url [1]{\endgroup\@href {#1}{\urlprefix }}%
\providecommand \urlprefix  [0]{URL }%
\providecommand \Eprint [0]{\href }%
\providecommand \doibase [0]{https://doi.org/}%
\providecommand \selectlanguage [0]{\@gobble}%
\providecommand \bibinfo  [0]{\@secondoftwo}%
\providecommand \bibfield  [0]{\@secondoftwo}%
\providecommand \translation [1]{[#1]}%
\providecommand \BibitemOpen [0]{}%
\providecommand \bibitemStop [0]{}%
\providecommand \bibitemNoStop [0]{.\EOS\space}%
\providecommand \EOS [0]{\spacefactor3000\relax}%
\providecommand \BibitemShut  [1]{\csname bibitem#1\endcsname}%
\let\auto@bib@innerbib\@empty
%</preamble>
\bibitem [{\citenamefont {Carter}(1993)}]{Carter}%
  \BibitemOpen
  \bibfield  {author} {\bibinfo {author} {\bibfnamefont {B.}~\bibnamefont
  {Carter}},\ }\bibfield  {title} {\bibinfo {title} {Perturbation dynamics for
  membranes and strings governed by the {D}irac-{N}ambu-{G}oto action in curved
  space},\ }\href@noop {} {\bibfield  {journal} {\bibinfo  {journal} {Phys.
  Rev. D}\ }\textbf {\bibinfo {volume} {48}} (\bibinfo {year}
  {1993})}\BibitemShut {NoStop}%
\bibitem [{\citenamefont {Carter}(2001)}]{Carterbrane}%
  \BibitemOpen
  \bibfield  {author} {\bibinfo {author} {\bibfnamefont {B.}~\bibnamefont
  {Carter}},\ }\bibfield  {title} {\bibinfo {title} {Essentials of classical
  brane dynamics},\ }\href@noop {} {\bibfield  {journal} {\bibinfo  {journal}
  {Int. J. Theor. Phys.}\ }\textbf {\bibinfo {volume} {40}} (\bibinfo {year}
  {2001})}\BibitemShut {NoStop}%
\bibitem [{\citenamefont {Zwiebach}(2009)}]{Zwiebach}%
  \BibitemOpen
  \bibfield  {author} {\bibinfo {author} {\bibfnamefont {B.}~\bibnamefont
  {Zwiebach}},\ }\href@noop {} {\emph {\bibinfo {title} {A First Course in
  String Theory}}},\ \bibinfo {edition} {2nd}\ ed.\ (\bibinfo  {publisher}
  {Cambridge University Press},\ \bibinfo {year} {2009})\BibitemShut {NoStop}%
\bibitem [{\citenamefont {Larsen}\ and\ \citenamefont {Frolov}(1994)}]{LF94}%
  \BibitemOpen
  \bibfield  {author} {\bibinfo {author} {\bibfnamefont {A.~L.}\ \bibnamefont
  {Larsen}}\ and\ \bibinfo {author} {\bibfnamefont {V.~P.}\ \bibnamefont
  {Frolov}},\ }\bibfield  {title} {\bibinfo {title} {Propagation of
  perturbations along strings},\ }\href@noop {} {\bibfield  {journal} {\bibinfo
   {journal} {Nucl. Phys. B}\ }\textbf {\bibinfo {volume} {414}},\ \bibinfo
  {pages} {129} (\bibinfo {year} {1994})}\BibitemShut {NoStop}%
\bibitem [{\citenamefont {Vilenkin}\ and\ \citenamefont
  {Shellard}(1994)}]{vilenkin1994cosmic}%
  \BibitemOpen
  \bibfield  {author} {\bibinfo {author} {\bibfnamefont {A.}~\bibnamefont
  {Vilenkin}}\ and\ \bibinfo {author} {\bibfnamefont {E.~P.~S.}\ \bibnamefont
  {Shellard}},\ }\href@noop {} {\emph {\bibinfo {title} {Cosmic strings and
  other topological defects}}}\ (\bibinfo  {publisher} {Cambridge University
  Press},\ \bibinfo {year} {1994})\BibitemShut {NoStop}%
\bibitem [{\citenamefont {Capovilla}\ and\ \citenamefont
  {Guven}(1995)}]{defos}%
  \BibitemOpen
  \bibfield  {author} {\bibinfo {author} {\bibfnamefont {R.}~\bibnamefont
  {Capovilla}}\ and\ \bibinfo {author} {\bibfnamefont {J.}~\bibnamefont
  {Guven}},\ }\bibfield  {title} {\bibinfo {title} {Geometry of deformations of
  relativistic membranes},\ }\href@noop {} {\bibfield  {journal} {\bibinfo
  {journal} {Phys. Rev. D}\ }\textbf {\bibinfo {volume} {51}} (\bibinfo {year}
  {1995})}\BibitemShut {NoStop}%
\bibitem [{\citenamefont {Ba$\dot{\mbox{z}}$a\'{n}ski}(1976)}]{Bazpol}%
  \BibitemOpen
  \bibfield  {author} {\bibinfo {author} {\bibfnamefont {S.~L.}\ \bibnamefont
  {Ba$\dot{\mbox{z}}$a\'{n}ski}},\ }\bibfield  {title} {\bibinfo {title}
  {Relative dynamics of the classical theory of fields},\ }\href@noop {}
  {\bibfield  {journal} {\bibinfo  {journal} {Acta Phys. Polon. B}\ }\textbf
  {\bibinfo {volume} {7}} (\bibinfo {year} {1976})}\BibitemShut {NoStop}%
\bibitem [{\citenamefont {Arreaga}\ \emph {et~al.}(2000)\citenamefont
  {Arreaga}, \citenamefont {Capovilla},\ and\ \citenamefont
  {Guven}}]{CapoNoether}%
  \BibitemOpen
  \bibfield  {author} {\bibinfo {author} {\bibfnamefont {G.}~\bibnamefont
  {Arreaga}}, \bibinfo {author} {\bibfnamefont {R.}~\bibnamefont {Capovilla}},\
  and\ \bibinfo {author} {\bibfnamefont {J.}~\bibnamefont {Guven}},\ }\bibfield
   {title} {\bibinfo {title} {Noether currents for bosonic branes},\
  }\href@noop {} {\bibfield  {journal} {\bibinfo  {journal} {Annals of
  Physics}\ }\textbf {\bibinfo {volume} {279}} (\bibinfo {year}
  {2000})}\BibitemShut {NoStop}%
\bibitem [{\citenamefont {Guven}(1993)}]{guven1993covariant}%
  \BibitemOpen
  \bibfield  {author} {\bibinfo {author} {\bibfnamefont {J.}~\bibnamefont
  {Guven}},\ }\bibfield  {title} {\bibinfo {title} {Covariant perturbations of
  domain walls in curved spacetime},\ }\href@noop {} {\bibfield  {journal}
  {\bibinfo  {journal} {Physical Review D}\ }\textbf {\bibinfo {volume} {48}},\
  \bibinfo {pages} {4604} (\bibinfo {year} {1993})}\BibitemShut {NoStop}%
\bibitem [{\citenamefont {Garriga}\ and\ \citenamefont
  {Vilenkin}(1991)}]{garriga1991perturbations}%
  \BibitemOpen
  \bibfield  {author} {\bibinfo {author} {\bibfnamefont {J.}~\bibnamefont
  {Garriga}}\ and\ \bibinfo {author} {\bibfnamefont {A.}~\bibnamefont
  {Vilenkin}},\ }\bibfield  {title} {\bibinfo {title} {Perturbations on domain
  walls and strings: A covariant theory},\ }\href@noop {} {\bibfield  {journal}
  {\bibinfo  {journal} {Physical Review D}\ }\textbf {\bibinfo {volume} {44}},\
  \bibinfo {pages} {1007} (\bibinfo {year} {1991})}\BibitemShut {NoStop}%
\bibitem [{\citenamefont {Battye}\ and\ \citenamefont {Carter}(2000)}]{BC00}%
  \BibitemOpen
  \bibfield  {author} {\bibinfo {author} {\bibfnamefont {R.~A.}\ \bibnamefont
  {Battye}}\ and\ \bibinfo {author} {\bibfnamefont {B.}~\bibnamefont
  {Carter}},\ }\bibfield  {title} {\bibinfo {title} {Second order lagrangian
  and symplectic current for gravitationally perturbed {D}irac-{N}ambu-{G}oto
  strings and branes},\ }\href@noop {} {\bibfield  {journal} {\bibinfo
  {journal} {Class. Quantum Grav.}\ }\textbf {\bibinfo {volume} {17}},\
  \bibinfo {pages} {3325} (\bibinfo {year} {2000})}\BibitemShut {NoStop}%
\bibitem [{\citenamefont {Mars}(2005)}]{Mars}%
  \BibitemOpen
  \bibfield  {author} {\bibinfo {author} {\bibfnamefont {M.}~\bibnamefont
  {Mars}},\ }\bibfield  {title} {\bibinfo {title} {First and second order
  perturbations of hypersurfaces},\ }\href@noop {} {\bibfield  {journal}
  {\bibinfo  {journal} {Class. Quantum Grav.}\ }\textbf {\bibinfo {volume}
  {22}},\ \bibinfo {pages} {3325} (\bibinfo {year} {2005})}\BibitemShut
  {NoStop}%
\bibitem [{\citenamefont {Bagatella-Flores}\ \emph {et~al.}(2016)\citenamefont
  {Bagatella-Flores}, \citenamefont {Campuzano}, \citenamefont {Cruz},\ and\
  \citenamefont {Rojas}}]{Lovelock}%
  \BibitemOpen
  \bibfield  {author} {\bibinfo {author} {\bibfnamefont {N.}~\bibnamefont
  {Bagatella-Flores}}, \bibinfo {author} {\bibfnamefont {C.}~\bibnamefont
  {Campuzano}}, \bibinfo {author} {\bibfnamefont {M.}~\bibnamefont {Cruz}},\
  and\ \bibinfo {author} {\bibfnamefont {E.}~\bibnamefont {Rojas}},\ }\bibfield
   {title} {\bibinfo {title} {Covariant approach of perturbations in {L}ovelock
  type brane gravity},\ }\href@noop {} {\bibfield  {journal} {\bibinfo
  {journal} {Class. Quantum Grav.}\ }\textbf {\bibinfo {volume} {33}} (\bibinfo
  {year} {2016})}\BibitemShut {NoStop}%
\bibitem [{\citenamefont {Fomenko}\ and\ \citenamefont {Tuzhilin}(2005)}]{FT}%
  \BibitemOpen
  \bibfield  {author} {\bibinfo {author} {\bibfnamefont {A.~T.}\ \bibnamefont
  {Fomenko}}\ and\ \bibinfo {author} {\bibfnamefont {A.~A.}\ \bibnamefont
  {Tuzhilin}},\ }\bibfield  {title} {\bibinfo {title} {Elements of the geometry
  and topology of minimal surfaces in {T}hree-{D}imensional {S}pace},\
  }\href@noop {} {\bibfield  {journal} {\bibinfo  {journal} {AMS, Translations
  of Mathematical Monographs}\ } (\bibinfo {year} {2005})}\BibitemShut
  {NoStop}%
\bibitem [{\citenamefont {Murorunkwere}\ \emph {et~al.}(2021)\citenamefont
  {Murorunkwere}, \citenamefont {Ntahompagaze},\ and\ \citenamefont
  {Jurua}}]{1+3fT}%
  \BibitemOpen
  \bibfield  {author} {\bibinfo {author} {\bibfnamefont {B.}~\bibnamefont
  {Murorunkwere}}, \bibinfo {author} {\bibfnamefont {J.}~\bibnamefont
  {Ntahompagaze}},\ and\ \bibinfo {author} {\bibfnamefont {E.}~\bibnamefont
  {Jurua}},\ }\bibfield  {title} {\bibinfo {title} {1 + 3 covariant
  perturbations in power-law f({R}) gravity},\ }\href@noop {} {\bibfield
  {journal} {\bibinfo  {journal} {Eur. Phys. J. C}\ }\textbf {\bibinfo {volume}
  {81}} (\bibinfo {year} {2021})}\BibitemShut {NoStop}%
\bibitem [{\citenamefont {Sahluu}\ \emph {et~al.}(2020)\citenamefont {Sahluu},
  \citenamefont {Ntahompagaze}, \citenamefont {Abebe}, \citenamefont {de~la
  Cruz-Dombriz},\ and\ \citenamefont {Mota}}]{1+3fR}%
  \BibitemOpen
  \bibfield  {author} {\bibinfo {author} {\bibfnamefont {S.}~\bibnamefont
  {Sahluu}}, \bibinfo {author} {\bibfnamefont {J.}~\bibnamefont
  {Ntahompagaze}}, \bibinfo {author} {\bibfnamefont {A.}~\bibnamefont {Abebe}},
  \bibinfo {author} {\bibfnamefont {I.}~\bibnamefont {de~la Cruz-Dombriz}},\
  and\ \bibinfo {author} {\bibfnamefont {D.~F.}\ \bibnamefont {Mota}},\
  }\bibfield  {title} {\bibinfo {title} {Scalar perturbations in f({T}) gravity
  using the 1 + 3 covariant approach},\ }\href@noop {} {\bibfield  {journal}
  {\bibinfo  {journal} {Eur. Phys. J. C}\ }\textbf {\bibinfo {volume} {80}}
  (\bibinfo {year} {2020})}\BibitemShut {NoStop}%
\bibitem [{\citenamefont {Kiosses}\ and\ \citenamefont
  {Nicolaidis}(2014)}]{KN14}%
  \BibitemOpen
  \bibfield  {author} {\bibinfo {author} {\bibfnamefont {V.}~\bibnamefont
  {Kiosses}}\ and\ \bibinfo {author} {\bibfnamefont {A.}~\bibnamefont
  {Nicolaidis}},\ }\bibfield  {title} {\bibinfo {title} {Second order
  perturbations of relativistic membranes in curved spacetime},\ }\href@noop {}
  {\bibfield  {journal} {\bibinfo  {journal} {Phys. Rev. D}\ }\textbf {\bibinfo
  {volume} {89}} (\bibinfo {year} {2014})}\BibitemShut {NoStop}%
\bibitem [{\citenamefont {Larsen}\ and\ \citenamefont
  {Nicolaidis}(2001)}]{LA01}%
  \BibitemOpen
  \bibfield  {author} {\bibinfo {author} {\bibfnamefont {A.~L.}\ \bibnamefont
  {Larsen}}\ and\ \bibinfo {author} {\bibfnamefont {A.}~\bibnamefont
  {Nicolaidis}},\ }\bibfield  {title} {\bibinfo {title} {Second order
  perturbations of a macroscopic string: Covariant approach},\ }\href@noop {}
  {\bibfield  {journal} {\bibinfo  {journal} {Phys. Rev. D}\ }\textbf {\bibinfo
  {volume} {63}} (\bibinfo {year} {2001})}\BibitemShut {NoStop}%
\bibitem [{\citenamefont {Ryu}\ and\ \citenamefont {Takayanagi}(2006)}]{RT06}%
  \BibitemOpen
  \bibfield  {author} {\bibinfo {author} {\bibfnamefont {S.}~\bibnamefont
  {Ryu}}\ and\ \bibinfo {author} {\bibfnamefont {T.}~\bibnamefont
  {Takayanagi}},\ }\bibfield  {title} {\bibinfo {title} {Aspects of holographic
  entanglement entropy},\ }\href@noop {} {\bibfield  {journal} {\bibinfo
  {journal} {JHEP}\ }\textbf {\bibinfo {volume} {08}}\bibinfo  {number} {
  (045)}}\BibitemShut {NoStop}%
\bibitem [{\citenamefont {Hubeny}\ \emph {et~al.}(2007)\citenamefont {Hubeny},
  \citenamefont {Rangamani},\ and\ \citenamefont {Takayanagi}}]{CovRT07}%
  \BibitemOpen
\bibfield  {number} {  }\bibfield  {author} {\bibinfo {author} {\bibfnamefont
  {V.~E.}\ \bibnamefont {Hubeny}}, \bibinfo {author} {\bibfnamefont
  {M.}~\bibnamefont {Rangamani}},\ and\ \bibinfo {author} {\bibfnamefont
  {T.}~\bibnamefont {Takayanagi}},\ }\bibfield  {title} {\bibinfo {title} {A
  covariant holographic entanglement entropy proposal},\ }\href@noop {}
  {\bibfield  {journal} {\bibinfo  {journal} {JHEP}\ }\textbf {\bibinfo
  {volume} {07}}\bibinfo  {number} { (062)}}\BibitemShut {NoStop}%
\bibitem [{\citenamefont {Bao}\ \emph {et~al.}(2019)\citenamefont {Bao},
  \citenamefont {Cao}, \citenamefont {Fischetti},\ and\ \citenamefont
  {Keeler}}]{Bao1}%
  \BibitemOpen
\bibfield  {number} {  }\bibfield  {author} {\bibinfo {author} {\bibfnamefont
  {N.}~\bibnamefont {Bao}}, \bibinfo {author} {\bibfnamefont {C.}~\bibnamefont
  {Cao}}, \bibinfo {author} {\bibfnamefont {S.}~\bibnamefont {Fischetti}},\
  and\ \bibinfo {author} {\bibfnamefont {C.}~\bibnamefont {Keeler}},\
  }\bibfield  {title} {\bibinfo {title} {Towards bulk metric reconstruction
  from extremal area variations},\ }\href@noop {} {\bibfield  {journal}
  {\bibinfo  {journal} {Class. Quantum Grav.}\ }\textbf {\bibinfo {volume}
  {36}} (\bibinfo {year} {2019})}\BibitemShut {NoStop}%
\bibitem [{\citenamefont {Bao}\ \emph {et~al.}(2021)\citenamefont {Bao},
  \citenamefont {Cao}, \citenamefont {Fischetti}, \citenamefont {Pollack},\
  and\ \citenamefont {Zhong}}]{Bao2}%
  \BibitemOpen
  \bibfield  {author} {\bibinfo {author} {\bibfnamefont {N.}~\bibnamefont
  {Bao}}, \bibinfo {author} {\bibfnamefont {C.}~\bibnamefont {Cao}}, \bibinfo
  {author} {\bibfnamefont {S.}~\bibnamefont {Fischetti}}, \bibinfo {author}
  {\bibfnamefont {J.}~\bibnamefont {Pollack}},\ and\ \bibinfo {author}
  {\bibfnamefont {Y.}~\bibnamefont {Zhong}},\ }\bibfield  {title} {\bibinfo
  {title} {More of the bulk from extremal area variations},\ }\href@noop {}
  {\bibfield  {journal} {\bibinfo  {journal} {Class. Quantum Grav.}\ }\textbf
  {\bibinfo {volume} {38}} (\bibinfo {year} {2021})}\BibitemShut {NoStop}%
\bibitem [{\citenamefont {Engelhardt}\ and\ \citenamefont
  {Fischetti}(2019)}]{netta}%
  \BibitemOpen
  \bibfield  {author} {\bibinfo {author} {\bibfnamefont {E.}~\bibnamefont
  {Engelhardt}}\ and\ \bibinfo {author} {\bibfnamefont {E.}~\bibnamefont
  {Fischetti}},\ }\bibfield  {title} {\bibinfo {title} {Surface theory: the
  classical, the quantum, and the holographic},\ }\href@noop {} {\bibfield
  {journal} {\bibinfo  {journal} {Class. Quantum Grav.}\ }\textbf {\bibinfo
  {volume} {36}} (\bibinfo {year} {2019})}\BibitemShut {NoStop}%
\bibitem [{\citenamefont {Martens}\ and\ \citenamefont {Koyama}(2010)}]{MK}%
  \BibitemOpen
  \bibfield  {author} {\bibinfo {author} {\bibfnamefont {R.}~\bibnamefont
  {Martens}}\ and\ \bibinfo {author} {\bibfnamefont {K.}~\bibnamefont
  {Koyama}},\ }\bibfield  {title} {\bibinfo {title} {Brane-world gravity},\
  }\href@noop {} {\bibfield  {journal} {\bibinfo  {journal} {Living Rev.
  Relativity}\ }\textbf {\bibinfo {volume} {13}},\ \bibinfo {pages} {5}
  (\bibinfo {year} {2010})}\BibitemShut {NoStop}%
\bibitem [{\citenamefont {Regge}\ and\ \citenamefont
  {Teitelboim}(1977)}]{regge2016general}%
  \BibitemOpen
  \bibfield  {author} {\bibinfo {author} {\bibfnamefont {T.}~\bibnamefont
  {Regge}}\ and\ \bibinfo {author} {\bibfnamefont {C.}~\bibnamefont
  {Teitelboim}},\ }\bibfield  {title} {\bibinfo {title} {General relativity
  \`{a} la string: a progress report},\ }\href@noop {} {\bibfield  {journal}
  {\bibinfo  {journal} {in \textit{Proceddings of the Marcel Grossman Meeting},
  Trieste, Italy (1975), ed. R. Ruffini (North-Holland, Amsterdam, 1977) 77
  (1977), arXiv:1612.05256}\ } (\bibinfo {year} {1977})}\BibitemShut {NoStop}%
\bibitem [{\citenamefont {Deserno}(2015)}]{Deserno}%
  \BibitemOpen
  \bibfield  {author} {\bibinfo {author} {\bibfnamefont {M.}~\bibnamefont
  {Deserno}},\ }\bibfield  {title} {\bibinfo {title} {Fluid lipid membranes:
  from differential geometry to curvature stresses},\ }\href@noop {} {\bibfield
   {journal} {\bibinfo  {journal} {Chemistry and Physics of Lipids}\ }\textbf
  {\bibinfo {volume} {185}},\ \bibinfo {pages} {11} (\bibinfo {year}
  {2015})}\BibitemShut {NoStop}%
\bibitem [{\citenamefont {Capovilla}(2018)}]{CapoSimLipid}%
  \BibitemOpen
  \bibfield  {author} {\bibinfo {author} {\bibfnamefont {R.}~\bibnamefont
  {Capovilla}},\ }\bibfield  {title} {\bibinfo {title} {A simultaneous
  variational principle for elementary excitations of fluid lipid membranes},\
  }\href@noop {} {\bibfield  {journal} {\bibinfo  {journal} {J. Phys. Comm.}\
  }\textbf {\bibinfo {volume} {2}} (\bibinfo {year} {2018})}\BibitemShut
  {NoStop}%
\bibitem [{\citenamefont {Capovilla}(2017)}]{CapoBending}%
  \BibitemOpen
  \bibfield  {author} {\bibinfo {author} {\bibfnamefont {R.}~\bibnamefont
  {Capovilla}},\ }\bibfield  {title} {\bibinfo {title} {Elastic bending energy:
  a variational approach},\ }\href@noop {} {\bibfield  {journal} {\bibinfo
  {journal} {Journal of Geometry and Symmetry in Physics}\ }\textbf {\bibinfo
  {volume} {45}},\ \bibinfo {pages} {1} (\bibinfo {year} {2017})}\BibitemShut
  {NoStop}%
\bibitem [{\citenamefont {Armas}\ \emph {et~al.}(2020)\citenamefont {Armas},
  \citenamefont {Hartong}, \citenamefont {Have}, \citenamefont {Nielsen},\ and\
  \citenamefont {Obers}}]{FluidMembranes20}%
  \BibitemOpen
  \bibfield  {author} {\bibinfo {author} {\bibfnamefont {J.}~\bibnamefont
  {Armas}}, \bibinfo {author} {\bibfnamefont {J.}~\bibnamefont {Hartong}},
  \bibinfo {author} {\bibfnamefont {E.}~\bibnamefont {Have}}, \bibinfo {author}
  {\bibfnamefont {B.~F.}\ \bibnamefont {Nielsen}},\ and\ \bibinfo {author}
  {\bibfnamefont {N.~A.}\ \bibnamefont {Obers}},\ }\bibfield  {title} {\bibinfo
  {title} {Newton-{C}artan submanifolds and fluid membranes},\ }\href@noop {}
  {\bibfield  {journal} {\bibinfo  {journal} {Phys. Rev. E}\ }\textbf {\bibinfo
  {volume} {101}} (\bibinfo {year} {2020})}\BibitemShut {NoStop}%
\bibitem [{\citenamefont {Armas}\ and\ \citenamefont
  {Tarr{\'\i}o}(2018)}]{Armas}%
  \BibitemOpen
  \bibfield  {author} {\bibinfo {author} {\bibfnamefont {J.}~\bibnamefont
  {Armas}}\ and\ \bibinfo {author} {\bibfnamefont {J.}~\bibnamefont
  {Tarr{\'\i}o}},\ }\bibfield  {title} {\bibinfo {title} {On actions for
  (entangling) surfaces and {DCFT}s},\ }\href@noop {} {\bibfield  {journal}
  {\bibinfo  {journal} {JHEP}\ }\textbf {\bibinfo {volume} {04}}\bibinfo
  {number} { (100)}}\BibitemShut {NoStop}%
\bibitem [{\citenamefont {Wald}(1984)}]{Wald}%
  \BibitemOpen
\bibfield  {number} {  }\bibfield  {author} {\bibinfo {author} {\bibfnamefont
  {R.}~\bibnamefont {Wald}},\ }\href@noop {} {\emph {\bibinfo {title} {General
  Relativity}}}\ (\bibinfo  {publisher} {U. Chicago Press},\ \bibinfo {year}
  {1984})\BibitemShut {NoStop}%
\bibitem [{\citenamefont {Gelfand}\ and\ \citenamefont {Fomin}(1963)}]{GF}%
  \BibitemOpen
  \bibfield  {author} {\bibinfo {author} {\bibfnamefont {I.~M.}\ \bibnamefont
  {Gelfand}}\ and\ \bibinfo {author} {\bibfnamefont {S.~V.}\ \bibnamefont
  {Fomin}},\ }\href@noop {} {\emph {\bibinfo {title} {Calculus of
  variations}}}\ (\bibinfo  {publisher} {Prentice-Hall},\ \bibinfo {year}
  {1963})\BibitemShut {NoStop}%
\bibitem [{\citenamefont {Kot}(2014)}]{Kot}%
  \BibitemOpen
  \bibfield  {author} {\bibinfo {author} {\bibfnamefont {M.}~\bibnamefont
  {Kot}},\ }\href@noop {} {\emph {\bibinfo {title} {A First Course in the
  Calculus of Variations}}},\ Vol.~\bibinfo {volume} {72}\ (\bibinfo
  {publisher} {AMS, Student Mathematical Library},\ \bibinfo {year}
  {2014})\BibitemShut {NoStop}%
\bibitem [{\citenamefont {Ba$\dot{\mbox{z}}$a\'{n}ski}(1977)}]{bz77b}%
  \BibitemOpen
  \bibfield  {author} {\bibinfo {author} {\bibfnamefont {S.~L.}\ \bibnamefont
  {Ba$\dot{\mbox{z}}$a\'{n}ski}},\ }\bibfield  {title} {\bibinfo {title}
  {Dynamics of relative motion of test particles in general relativity},\
  }\href@noop {} {\bibfield  {journal} {\bibinfo  {journal} {Ann. Inst. H.
  Poincar\'e A}\ }\textbf {\bibinfo {volume} {27}},\ \bibinfo {pages} {145}
  (\bibinfo {year} {1977})}\BibitemShut {NoStop}%
\bibitem [{\citenamefont {Cruz}\ and\ \citenamefont {Rojas}(2013)}]{CruzE}%
  \BibitemOpen
  \bibfield  {author} {\bibinfo {author} {\bibfnamefont {M.}~\bibnamefont
  {Cruz}}\ and\ \bibinfo {author} {\bibfnamefont {E.}~\bibnamefont {Rojas}},\
  }\bibfield  {title} {\bibinfo {title} {Born--infeld extension of {L}ovelock
  brane gravity},\ }\href@noop {} {\bibfield  {journal} {\bibinfo  {journal}
  {Class. Quantum Grav.}\ }\textbf {\bibinfo {volume} {30}} (\bibinfo {year}
  {2013})}\BibitemShut {NoStop}%
\bibitem [{\citenamefont {Ba$\dot{\mbox{z}}$a\'{n}ski}(1989)}]{bz89}%
  \BibitemOpen
  \bibfield  {author} {\bibinfo {author} {\bibfnamefont {S.~L.}\ \bibnamefont
  {Ba$\dot{\mbox{z}}$a\'{n}ski}},\ }\bibfield  {title} {\bibinfo {title}
  {Hamilton--{J}acobi formalism for geodesics and geodesic deviations},\
  }\href@noop {} {\bibfield  {journal} {\bibinfo  {journal} {J. Math. Phys.}\
  }\textbf {\bibinfo {volume} {30}} (\bibinfo {year} {1989})}\BibitemShut
  {NoStop}%
\bibitem [{\citenamefont {Capovilla}\ and\ \citenamefont {Cruz}(2019)}]{CG1}%
  \BibitemOpen
  \bibfield  {author} {\bibinfo {author} {\bibfnamefont {R.}~\bibnamefont
  {Capovilla}}\ and\ \bibinfo {author} {\bibfnamefont {G.}~\bibnamefont
  {Cruz}},\ }\bibfield  {title} {\bibinfo {title} {A covariant simultaneous
  action for branes},\ }\href@noop {} {\bibfield  {journal} {\bibinfo
  {journal} {Annals of Physics}\ }\textbf {\bibinfo {volume} {411}} (\bibinfo
  {year} {2019})}\BibitemShut {NoStop}%
\bibitem [{\citenamefont {Giaquinta}\ and\ \citenamefont
  {Hildebrandt}(2004)}]{GH}%
  \BibitemOpen
  \bibfield  {author} {\bibinfo {author} {\bibfnamefont {M.}~\bibnamefont
  {Giaquinta}}\ and\ \bibinfo {author} {\bibfnamefont {S.}~\bibnamefont
  {Hildebrandt}},\ }\href@noop {} {\emph {\bibinfo {title} {Calculus of
  variations}}}\ (\bibinfo  {publisher} {Springer},\ \bibinfo {year}
  {2004})\BibitemShut {NoStop}%
\bibitem [{\citenamefont {L{\'o}pez}(2018)}]{Edgarthesis}%
  \BibitemOpen
  \bibfield  {author} {\bibinfo {author} {\bibfnamefont {E.~Y.}\ \bibnamefont
  {L{\'o}pez}},\ }\href@noop {} {\emph {\bibinfo {title} {Perturbaciones de
  segundo orden de Branas}}},\ \bibinfo {organization} {M.Sc. Thesis,
  CINVESTAV-IPN} (\bibinfo {year} {2018})\BibitemShut {NoStop}%
\bibitem [{\citenamefont {Dirac}(1962)}]{dirac1962extensible}%
  \BibitemOpen
  \bibfield  {author} {\bibinfo {author} {\bibfnamefont {P.~A.~M.}\
  \bibnamefont {Dirac}},\ }\bibfield  {title} {\bibinfo {title} {An extensible
  model of the electron},\ }\href@noop {} {\bibfield  {journal} {\bibinfo
  {journal} {Proceedings of the Royal Society of London. Series A. Mathematical
  and Physical Sciences}\ }\textbf {\bibinfo {volume} {268}},\ \bibinfo {pages}
  {57} (\bibinfo {year} {1962})}\BibitemShut {NoStop}%
\bibitem [{\citenamefont {Nambu}(1995)}]{nambu1970duality}%
  \BibitemOpen
  \bibfield  {author} {\bibinfo {author} {\bibfnamefont {Y.}~\bibnamefont
  {Nambu}},\ }\bibfield  {title} {\bibinfo {title} {Duality and
  {H}adrodynamics; lectures at the 1970 {C}openhagen {H}igh {E}nergy {S}ymp},\
  }in\ \href@noop {} {\emph {\bibinfo {booktitle} {Broken Symmetry: Selected
  papers of {Y}. {N}ambu}}},\ \bibinfo {editor} {edited by\ \bibinfo {editor}
  {\bibfnamefont {T.}~\bibnamefont {Eguchi}}\ and\ \bibinfo {editor}
  {\bibfnamefont {K.}~\bibnamefont {Nishijima}}}\ (\bibinfo  {publisher} {World
  Scientific},\ \bibinfo {year} {1995})\ pp.\ \bibinfo {pages}
  {280--301}\BibitemShut {NoStop}%
\bibitem [{\citenamefont {Got{\=o}}(1971)}]{goto1971relativistic}%
  \BibitemOpen
  \bibfield  {author} {\bibinfo {author} {\bibfnamefont {T.}~\bibnamefont
  {Got{\=o}}},\ }\bibfield  {title} {\bibinfo {title} {Relativistic quantum
  mechanics of one-dimensional mechanical continuum and subsidiary condition of
  dual resonance model},\ }\href@noop {} {\bibfield  {journal} {\bibinfo
  {journal} {Progress of Theoretical Physics}\ }\textbf {\bibinfo {volume}
  {46}},\ \bibinfo {pages} {1560} (\bibinfo {year} {1971})}\BibitemShut
  {NoStop}%
\end{thebibliography}%
\end{document}